# Comment: Disappearance of Metal-Like Behavior in GaAs Two-Dimensional Holes below 30 mK

Huang et al.[1] have reported the resistivity T-dependence of holes in GaAs (HIGFET) for a T-range 150mK to 0.5mK at carrier densities ($p>p_c$) for which metallic behavior has been observed. They report a minimum in $\rho(T)$ at $T_m \sim$ 25mK for $p \sim 7.4 \times 10^9/cm^2$ and $T_m \sim 32$mK for $p \sim 8 \times 10^9/cm^2$. In each instance $\rho(T)$ slowly rises for $T<T_m$ to a value less than 10% above $\rho(T_m)$ as $T \to 0$. Huang et. al. report a lower density ($p \sim 5.2 \times 10^9/cm^2$, but $p>p_c$) where no minimum is seen down to 1mK. These authors suggest the minimum and the change from $d\rho/dT>0$ (normal metallic behavior) for $T>T_m$ to $d\rho/dT<0$ for $T<T_m$ indicates a change to insulator-like behavior. Such an interpretation is clearly relevant to whet-her the extensive 2D experimental studies indi-cating a 2D quantum phase transition (QTP) and whether metallic behavior survives as $T \to 0K$. This Comment suggests the data is explained by a 2D Kondo[2] effect and doesn't represent a crossover from metallic to insulating behavior.

The resistance minimum has a long history starting with the noble metals[3]. Franck et al.[4] for Cu:Fe demonstrated a minimum in $\rho(T)$ at 36K (0.2%, 31.5K (0.1%), and 25.5K (0.05%Fe). This rise in $\rho(T)$ below a $T_m$ was recognized as due to carrier scattering by local moments (Fe in Cu) and became known as the Kondo effect after a theory of the rise in $\rho(T)$ for $T<T_m$. The Kondo temperature $T_K = [\langle E \rangle /k] \exp(-1/N(E_F)|J|)$ where $\langle E \rangle$ is a characteristic energy, $N(E_F)$ is the density-of-states (DOS) and J is the mean exchange interaction (antiferromagnetic, J<0) between the localized moments and the carriers. In the 3D case the $T_K$ decreases as the local moment concentration is decreased. A change in Fe from 0.2% to 0.005% only changes $T_K$ from 36K to 17K. Ionov and Shlimak[5] observed the Kondo effect in degenerate p-type Ge ($p \sim 1.6 \times 10^{18}/cm^3$) and found $T_m \sim 6K$ and $\rho(0)/\rho(T_m) \sim 1.01$.

The 2D results in [1] show $T_m$ dropping from 32mK to less than 1mK. This large change in $T_m$ ($T_K$), where the concentration of local moments is presumed constant (independent of p and T) can be explained by a drop in $N(E_F)$ as $p \to p_c$ resulting from the strong interactions and the inter-action-induced pseudogap. For 2D it has been shown[6] $N(E_F) = a_2 [k_F a^*/(A_i + k_F a^*)] = a_2 \alpha(x)$ where $a_2 = m_h/\pi \hbar^2$ [$a_2$ noninteracting DOS, $m_h$ the hole effective mass], $a^*$ the acceptor Bohr radius and $A_i$ a measure of the net interaction strength, and $k_F = (2\pi p_c x)^{1/2}$ [$x=(p/p_c-1)$, $p_c \sim 4.2 \times 10^9/cm^2$]. As p is reduced from $8 \times 10^9$ to $5.2 \times 10^9/cm^2$ $N(E_F)$ is reduced by a factor of 1.8 and $1/N(E_F)|J|$ increases by the same amount for a fixed J. For $p \sim 8 \times 10^9/cm^2$ the data is fit to the $T_K$ expression to determine $a_2|J|$.

Table: Parameters for $T_K$

| p $\times 10^9$ | x | $T_F$(K) | $\alpha(x)$ | $T_K$ (mK) |
|---|---|---|---|---|
| 5.2 | .238 | 1.2 | 0.112 | 0.566 |
| 7.4 | .762 | 2.25 | 0.185 | 21.3 |
| 8.0 | .905 | 2.47 | 0.198 | 32 fit |

based on $p_c=4.2 \times 10^9/cm^2$; $A_i=0.27$; $a_2|J|= 1.163$

For $p \sim 5.2 \times 10^9/cm^2$ $T_K$ is reduced by a factor of 57 thus explaining why no minimum is seen for this more dilute metallic sample. The identity and concentration of the local moments is not known and is not necessarily from a 3d or 4f element impurity. Hamilton et. al.[7] also reported a $\rho(T)$ minimum in 2D p-type GaAs with $T_m \sim 0.22K$ and $\rho(0)/\rho(T_m) \sim 1.004$ for $p \sim 1.4 \times 10^{11}/cm^2$ where $p_c \sim 0.73 \times 10^{11}/cm^2$. This $p_c$ is 17 times that in [1] and $N(E_F)$ will be larger by a factor of 4. The 6.9 factor increase in $T_m$ is not surprising. The local moment concentration probably differs from that in [1]. The results in [1] showing a low-T min-imum in $\rho(T)$ exhibit the characteristics of a 2D Kondo effect and therefore indicate metallic be-havior to 0.5mK. This data provides new strong support for a 2D QPT as $T \rightarrow 0$. To establish insulating behavior as $T \rightarrow 0$ one needs to show that $\rho(T \rightarrow 0) \rightarrow \infty$ and that $|d\rho/dT|$ gets larger and larger as $T \rightarrow 0$. A large body of data show this is only the case for $p < p_c$.

T.G. Castner

Department of Physics and Astronomy, University of Rochester, Rochester, NY 14627